\documentclass[useAMS]{mn2e}
\usepackage{graphicx}
\usepackage{epsfig}
\usepackage{amssymb}
\usepackage{lscape}
\usepackage{ulem}
\usepackage{txfonts}

\def\kms{km~s$^{-1}$}

\def\c2s{C\,{\sc ii}$^{\star}$}
\def\hkpc{$h_{70}^{-1}$ kpc}

\def\dv{$\Delta$V}

\def\rp{$r_p$}

\def\fy{f$_{\rm young}$}

\title[The properties of post-merger galaxies] {Galaxy pairs in the Sloan 
Digital Sky Survey - VIII: The observational properties of post-merger galaxies.}

\author[Ellison et al.] {Sara L. Ellison$^1$,
J. Trevor Mendel$^2$,
David R. Patton$^3$,
Jillian M. Scudder$^1$\\ 
$^1$ Department of Physics \& Astronomy, University
of Victoria, Finnerty Road, Victoria, British Columbia, V8P 1A1,
Canada.\\
$^2$ Max-Planck-Institut fur Extraterrestrische Physik, Giessenbachstrasse, D-85748 Garching, Germany.\\
$^3$ Department of Physics \& Astronomy, Trent University,
1600 West Bank Drive, Peterborough, Ontario, K9J 7B8, Canada.
}
\begin{document}

\maketitle

\begin{abstract}

In order to investigate the effects of galaxy mergers throughout the
interaction sequence, we present a study of 10,800 galaxies in close
pairs and a smaller sample of 97 post-mergers identified in the Sloan
Digital Sky Survey.  We find that the average central star formation
rate (SFR) enhancement ($\times 3.5$) and the fraction of starbursts
(20 per cent) peak in the post-merger sample.  The post-mergers also
show a stronger deficit in gas phase metallicity than the closest
pairs, being more metal-poor than their control by $-0.09$ dex.
Combined with the observed trends in SFR and the timescales predicted
in merger simulations, we estimate that the post-mergers in our sample
have undergone coalescence within the last few hundred Myr.  In
contrast with the incidence of star-forming galaxies, the frequency of
active galactic nuclei (AGN) peaks in the post-mergers, outnumbering
AGN in the control sample by a factor of 3.75.  Moreover, amongst the
galaxies that host an AGN, the black hole accretion rates in the
closest pairs and post-mergers are higher by a factor of $\sim$ 3 than
AGN in the control sample.  These results are consistent with a
picture in which star formation is initiated early on in the
encounter, with AGN activity peaking post-coalescence.

\end{abstract}

\begin{keywords}
Galaxies: interactions, galaxies: abundances, galaxies: active, galaxies:
evolution, galaxies: Seyfert, galaxies: starburst
\end{keywords}

\section{Introduction}

Galaxy mergers represent a cornerstone in our picture of hierarchical
galaxy evolution.  In addition to building up the stellar mass in
galaxies, mergers are predicted to trigger central starbursts, feed
the central supermassive black hole and potentially completely
transform the galaxy's morphology (e.g. Barnes \& Hernquist 1996; Di
Matteo, Springel \& Hernquist 2005; Springel, Di Matteo \& Hernquist
2005; Di Matteo et al. 2007, 2008; Cox et al. 2008; Montuori et
al. 2010; Rupke, Kewley \& Barnes 2010; Perez et al. 2011; Torrey et
al. 2012).  There has been considerable observational work over the
last decades to confirm these theoretical predictions, which bear out
many of the general, qualitative predictions of the simulations over a
range of redshifts (e.g. Kennicutt et al. 1987; Barton, Geller \&
Kenyon 2000; Kewley, Geller \& Barton 2006; Woods, Geller \& Barton
2006; Smith et al.  2007; Jogee et al. 2009; Rupke, Kewley \& Chien
2010; Koss et al. 2010, 2012; Silverman et al. 2011; Wong et al. 2011;
Ramos-Almeida et al. 2011, 2012; Xu et al. 2012; Cotini et al. 2013;
Lanz et al. 2013).  However, a full and detailed picture of the merger
transformation requires large samples of galaxies that span a range in
mass, mass ratio, structural properties, environments and projected
separation, since all of these properties are likely to impact the
specific outcome of a merger.  Moreover, a combination of both
spectroscopic and imaging data, which yields simultaneous information
on morphology, metallicity, star formation and black hole accretion
rates etc. is needed to fully characterize the changes that these
galaxies undergo.  Finally, a carefully matched control sample is
essential in order to disentangle the effects of mergers from trends
that are intrinsically present in the galaxy population (Perez et
al. 2009).

The Sloan Digital Sky Survey (SDSS) is well suited to this task,
providing public imaging and spectroscopic data for approximately one
million galaxies in the nearby universe.  By combining the imaging and
spectroscopic samples, it has been possible to tackle many aspects of
galaxy mergers with unparalleled statistics.  Samples of either close
spectroscopic pairs, or morphologically classified mergers have been
used to investigate the impact of galaxy interactions on colours
(Ellison et al. 2010; Patton et al. 2011; Alonso et al. 2012; Lambas
et al. 2012), star formation rates (Nikolic, Cullen \& Alexander 2004;
Alonso et al. 2006; Ellison et al. 2008, 2013; Li et al 2008a; Scudder
et al. 2012b; Patton et al. 2013), metallicity (Michel-Dansac et
al. 2008; Ellison et al. 2008; Reichard et al. 2009; Peeples et
al. 2009; Alonso, Michel-Dansac \& Lambas 2010; Scudder et al. 2012b;
Chung et al. 2013), morphology (Ellison et al.  2010; Darg et
al. 2010; Casteels et al. 2013), gas consumption (Fertig et al.  in
preparation; Scudder et al. in preparation), infra-red emission (Hwang
et al. 2010, 2011; Ellison et al. 2013) and nuclear activity (Alonso
et al. 2007; Woods \& Geller 2007; Li et al. 2008b; Ellison et
al. 2008, 2011; Rogers et al. 2009; Darg et al. 2010; Liu, Shen \&
Strauss 2012; Sabater et al. 2013).

Although the observational signatures of mergers are now well
documented, quantifying the relevent timescales for star formation and
black hole accretion is more challenging to pin down.  Early
simulations (e.g. Mihos \& Hernquist 1994, 1996) found that the
duration of merger-triggered starbursts was fairly short, on the order
of tens of megayears.  At the time, similarly short timescales had
also been inferred from observations of mergers (e.g. Larson \&
Tinsley 1978; Kennicutt et al. 1987).  The natural projected
separations on which to study the effects of mergers would therefore
be at most a few tens of kpc.  In more recent works, the
observationally informed duration of triggered star formation has been
extended to several hundred megayears (e.g. Barton, Geller \& Kenyon
2000; Knapen \& James 2009; Woods et al. 2010).  For relative
velocities on the order of a few hundred \kms, this corresponds to
projected separations up to $\sim$ 50 \hkpc.  Indeed, this has been the
typical separation that most studies of close pairs have selected,
with enhanced star formation rates (SFRs) usually detected within
projected separations $\sim$ 30 \hkpc\ (e.g. Barton et al. 2000;
Alonso et al.  2007; Domingue et al. 2009; Wong et al. 2011). However,
it has recently been shown that the effects of galaxy interactions can
be detected in pairs with separations in excess of 50 \hkpc\ 
(e.g. Park \& Choi 2009; Scudder
et al. 2012b; Patton et al.  2011). Using a new methodology for 
identifying pairs and control samples out to projected separations of 
1 Mpc, Patton et
al. (2013) have recently shown that the SFRs in SDSS galaxy pairs only
decline to the level of a control sample at separations beyond 150 \hkpc.
Perhaps even more surprisingly, Patton et al. (2013) also determined
that the integrated star formation rate enhancement in galaxy
interactions is actually \textit{dominated} by widely separated pairs,
($50 < r_p < 150$ \hkpc) which are often excluded from merger studies.
These results appear to be robust, and not simply a feature of the
analysis technique, as has recently been shown to apply to correlation
function analyses (Robaina \& Bell 2012).  Moreover, the detection of
enhanced SFRs at wide separations is consistent with the much longer
timescales (up to a Gyr or more) for interaction-induced star
formation predicted by contemporary hydrodynamical simulations
(e.g. Cox et al 2006, 2008; Torrey et al 2012).  The short timescales
inferred from early observational studies (e.g. Larson
\& Tinsley 1978) may be in part a selection
bias whereby the morphological disturbances (as a sign of interaction)
favour short post-pericentre timescales, whereas samples of pairs more
fully sample the full interaction sequence (Lotz et al. 2008).

With the recent extension of merger samples out to projected
separations of 1 Mpc, and with the interpretive framework provided by
modern simulations, we are now converging on a fairly complete view of
galaxy-galaxy interactions.  However, there is one piece of the merger
story that is still missing from studies of galaxy pairs: final
coalescence and post-merger evolution.  Merger simulations (e.g. Mihos
\& Hernquist 1996; Di Matteo et al 2005; Springel et al. 2005; Cox et
al. 2006; Di Matteo et al. 2007; Johansson, Naab \& Burkert 2009;
Montuori et al. 2010; Debuhr, Quataert \& Ma 2011;
Torrey et al. 2012) indicate that the most
intense bursts of star formation (and black hole accretion) happen
around the time of coalescence, so it is important to include this
component in observational studies.  Certain selection criteria are
known to frequently identify late stage mergers, most notably the
luminous and ultra-luminous infra-red galaxies (LIRGs and ULIRGs, e.g.
Kartaltepe et al. 2010 and references therein).  However, selection
based on a fixed IR luminosity is likely to be biased towards the most
massive, most highly star-forming mergers (e.g. Hwang et al. 2010;
Ellison et al. 2013).  Very few systematic studies of the post-merger
phase exist.  Carpineti et al. (2012) identify a sample of 30
spheroidal post-mergers from which they measure an enhanced fraction
of active galactic nuclei (AGN).  However, without a well matched and
homogeneously analysed sample of galaxies in the pre-merger phase, it
is not straightforward to quantify the impact of final coalescence
relative to earlier stages in the interaction sequence.

The goal of this paper is to identify a sample of post-merger galaxies
in the SDSS and analyse them identically to the sample of close pairs
studied in our previous works.  In this way, it is possible to follow
the evolution of the merger to its final stages.  One of the strengths
of this work is that we can apply identical analysis techniques to the
pre- and post-merger phases, ensuring that the data are comparable in
the two regimes.  In Section \ref{sample_sec} we describe the various
samples used in this work: close pairs, post-mergers and their
respective controls. Sections \ref{sfr_sec}, \ref{metal_sec} and
\ref{agn_sec} deal with changes in the star-forming, metallicity and
nuclear properties, respectively.

We adopt a cosmology of  $\Omega_{\Lambda} = 0.7$, $\Omega_M = 0.3$,
$H_0 = 70$ km/s/Mpc.

\section{Sample selection}\label{sample_sec}

\subsection{Close pairs}

The starting point for the selection of our sample of galaxy pairs is
the SDSS DR7 Main Galaxy Sample (14.0 $\le m_r \le$ 17.77) with a
redshift range $0.01 \le z \le 0.2$ and SDSS specclass=2.  The
criteria and justification for our selection of pairs from the SDSS
has been discussed extensively in our previous papers on this topic
(e.g. Patton et al. 2011; Scudder et al. 2012b; Ellison et al. 2011,
2013).  We therefore review only briefly the main requirements and
refer the interested reader to these earlier works for more complete
details and discussions on selection biases etc.  In order to be
included in our sample of close pairs, we require that a galaxy has a
companion that fulfills the following criteria:

\begin{enumerate}

\item  A projected physical separation \rp\ $\le$ 80 \hkpc.

\item  A line of sight velocity difference \dv\ $\le$300 \kms, in order
to minimize chance projections.

\item A stellar mass ratio of 0.25 $\le$ M$_1$/M$_2$ $\le$ 4.  

\end{enumerate}

There are two main differences in the pairs sample used in this paper
compared to our previous works.  The first is our choice of mass
ratios; in our previous papers, we have investigated the effect of
mergers on pairs with stellar masses that differ by up to a
factor of 10.  However, the goal of the current study is to investigate the
evolution of mergers as they progress through the coalescence phase.
It is therefore desirable to identify close pairs that are, as far as
is possible, pre-merger analogs to the post-mergers in our sample.
The visual or automated selection of post-mergers, both of which
rely on galaxy asymmetry, is likely to be 
highly skewed towards major mergers, with approximately equal stellar 
masses (e.g. Lotz et al. 2010).
We have therefore restricted our sample to pairs whose mass ratios are
within a factor of 4 of one another.  However, it is worth noting that
our results change very little with this more restrictive mass ratio
cut, as may have been expected based on our previous work
(e.g. Ellison et al. 2011; Scudder et al. 2012b).  This invariance
also means that our exact choice of mass ratio criterion does
not alter any of our results.

\begin{figure}
\centerline{\rotatebox{270}{\resizebox{8cm}{!}
{\includegraphics{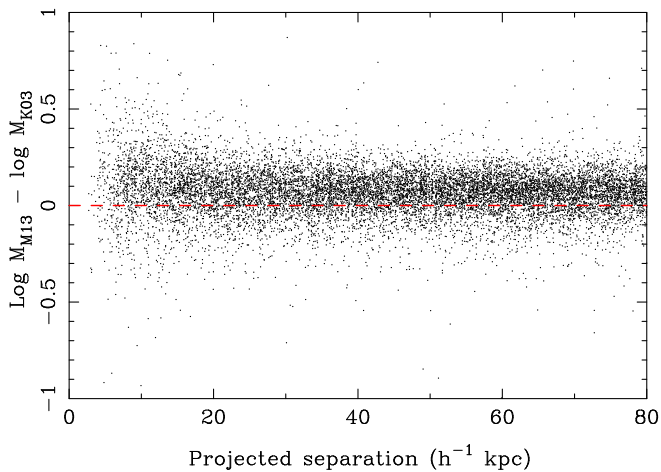}}}}
\caption{\label{mass} Comparison between the Mendel et al. (2013a) and
Kauffmann et al. (2003a) stellar masses (denoted as M13 and K03, respectively,
in the figure) for our sample of close pairs, as a function of
projected separation.  }
\end{figure}

The second main difference in the work presented here compared with
our previous pairs' studies, is that we adopt total stellar masses
determined from the Sersic fits presented by Mendel et al. (2013a),
rather than those derived by Kauffmann et al. (2003a).  The Mendel et
al. (2013a) masses use the updated SDSS photometry provided by Simard
et al. (2011), which has been demonstrated to be robust even in the
presence of a very close companion.  As shown by Mendel et
al. (2013a), the two sets of masses are generally in good agreement,
although the masses of Kauffmann et al. (2003a) are, $\sim$ 0.05 --
0.1 dex lighter than those derived by Mendel et al. (2013a), depending
on the mass.  In Figure \ref{mass} we compare the Mendel (2013a) and
Kauffmann et al. (2003a) masses as a function of projected separation.
In addition to the slight offset to higher masses in the Mendel et
al. (2013a) catalog, it can be seen that not only is the scatter
between the two values larger at the smallest separations, but there
is a an overall increase in the trend towards lighter masses in the
Kauffmann et al. (2003a) values at projected separations \rp\ $<$ 20
\hkpc. This is due to the use of improved photometry in crowded
environments by Mendel et al. (2013a) for their mass
determinations. For the post-mergers, we find that the masses from the
two catalogs are in good agreement.

The SDSS is subject to significant spectroscopic incompleteness at
angular separations $\theta < 55$ arcsec (Patton \& Atfield 2008).  If
this incompleteness is not accounted for, a significant selection bias
is introduced (e.g. Ellison et al. 2008). We therefore follow the
strategy of our earlier works and apply a random culling\footnote{An 
alternative
strategy, which we have applied in Patton et al. (2013), is to apply a
statistical weight that is a function of the projected separation.
Since Patton et al. (2013) were particularly interested in wide
separation pairs, assigning a higher weight to galaxies with small
separations was preferable to simply excluding two-thirds of the
sample at wide separations.} of 67.5\% of
pairs at separations $\theta > 55$ arcsec.  There are 10,800 galaxies 
in the close pairs sample after the culling has been applied.

As described in the Introduction, our previous work has demonstrated
that pairs remain significantly different from their control samples
out to projected separations of $\sim$ 150 \hkpc\ (Patton et
al. 2013).  However, in the work presented here, we do not include the
full wide pairs sample, primarily because it requires a more complex
control matching procedure that is not necessary for this work.  We
note that, as shown in Patton et al. (2013), SFR enhancements of the
close pairs follow a gradual decline out to the wider separations; in
this paper, we focus on extending the investigation of properties to
the final stage of the merger sequence: coalescence.

\subsection{Post-mergers}

In order to identify galaxies in the final stages of their
interaction, after the two individual galaxies have coalesced, we turn
to the visual classifications of SDSS galaxies performed by the Galazy
Zoo project.  The basic Galazy Zoo sample
selection includes $\sim$ 900,000 galaxies from the SDSS DR6, as
described by Lintott et al (2008).  Darg et al. (2010) impose further
requirements that the spectroscopic redshift of a galaxy must be 0.005
$< z <$ 0.1 and the extinction corrected $r$-band Petrosian magnitude
be $m_r < 17.77$, from which they construct samples of galaxy mergers
\footnote{In addition to the basic morphological classifications of
`spiral' and `elliptical', Galaxy Zoo volunteers may also identify
galaxies as `mergers'.}.  Based on an initial cut in the total merger
vote fractions ($f_m > 0.4$), followed by further visual follow-up by
a professional astronomer, Darg et al. (2010) have compiled two
distinct Galazy Zoo merger catalogues.  The first consists of binary
mergers, akin to a morphologically selected version of our own
spectroscopic close pairs sample.  The second catalogue consists of
post-mergers; single galaxies that are strongly disturbed, highly
irregular, exhibit significant tidal features, or other signs of a
recent interaction.  The post-merger catalogue consists of 358
galaxies.  By construction, the post-merger catalogue will only
include those merger events that are sufficient to produce strong
tidal features or otherwise visibly distort the galaxies in the SDSS
images.  

\begin{figure}
\centerline{\rotatebox{0}{\resizebox{9cm}{!}
{\includegraphics{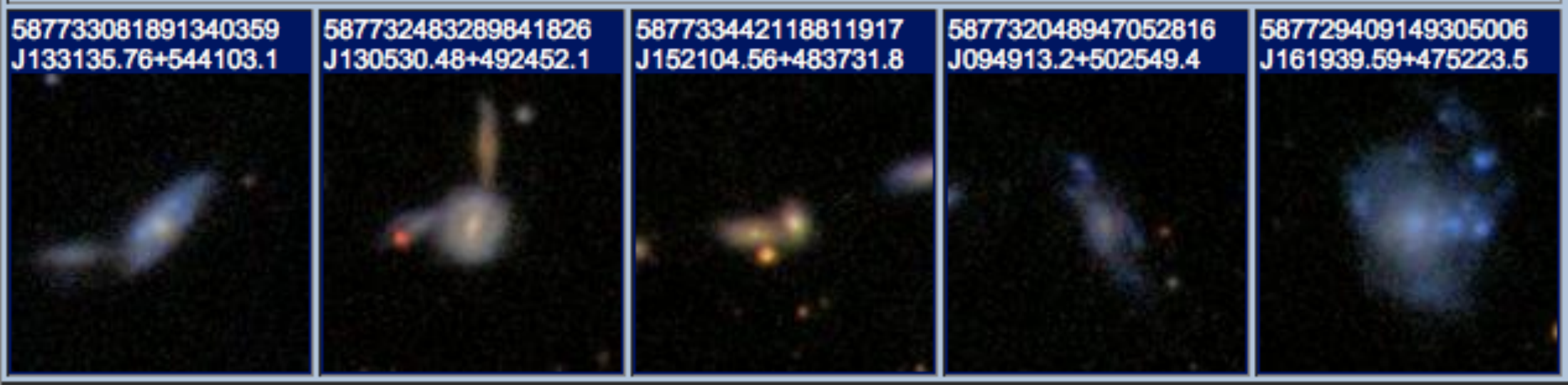}}}}
\caption{\label{reject} Examples of galaxies originally classified as
post-mergers in the sample of Darg et al. (2010), but rejected during
our visual classifications. The first 3 have a 
resolved companion, the last 2 are classified as `normal' irregular
galaxies. The SDSS object ID is given in the top of each panel.}
\end{figure}

\begin{figure}
\centerline{\rotatebox{0}{\resizebox{9cm}{!}
{\includegraphics{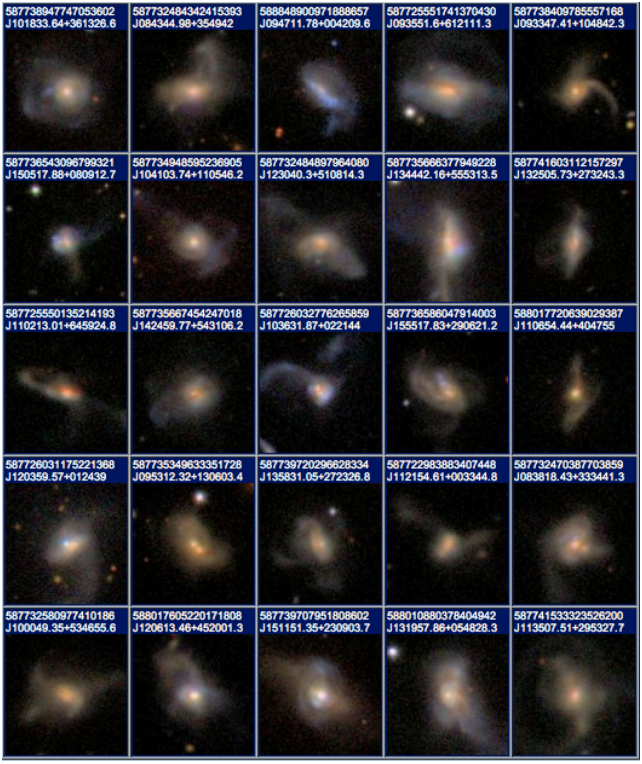}}}}
\caption{\label{pm_gals} A selection of 25 post-mergers from our
sample of 97 that remain after our own visual classification of the
Galaxy Zoo catalog. The SDSS object ID is given in the top of each panel.}
\end{figure}

We have performed a further visual classification of galaxies
in the post-merger catalogue in order to remove either a) galaxies
that are irregular, but not apparently merging; these galaxies
are consistent with morphologies of similar
galaxies in the Hubble classification or b) late phase encounters 
(close pairs) that
have not fully coalesced to a final post-merger state.  Examples of
these rejected categories are shown in Figure \ref{reject}.  We also
require that, to be consistent with the selection of the close pairs,
the post-merger galaxies have an SDSS specclass=2 and a Mendel et
al. (2013a) mass available.  The final sample of post-mergers consists
of 97 galaxies, a selection of which is shown in Figure
\ref{pm_gals}.

% To make these - first do a screen grab on mac, then convert to
% pdf in preview, then transfer here.

\begin{figure}
\centerline{\rotatebox{0}{\resizebox{9cm}{!}
{\includegraphics{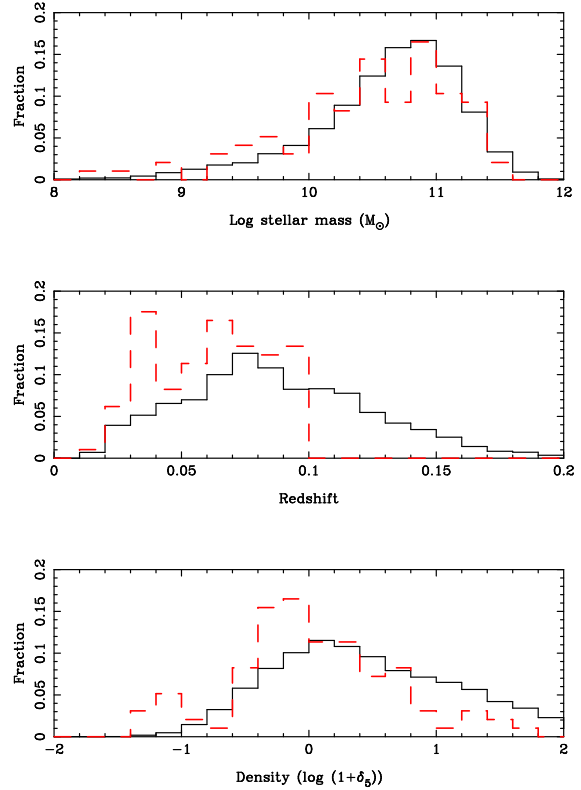}}}}
\caption{\label{mass_z} The distribution of stellar masses (upper
panel), redshifts (middle panel) and local overdensities (lower panel)
for the close pairs (black solid line) and post-mergers (red dashed
line).}
\end{figure}

\medskip

In Figure \ref{mass_z} we show the distribution of stellar masses,
redshifts and local overdensity for both the post-merger and close
pairs samples.  Following our previous works, we compute local
environmental densities, $\Sigma_n$:

\begin{equation}
\Sigma_n = \frac{n}{\pi d_n^2},
\end{equation}

where $d_n$ is the projected distance in Mpc to the $n^{th}$ nearest
neighbour within $\pm$1000 \kms.  Normalized densities, $\delta_n$,
are computed relative to the median $\Sigma_n$ within a redshift slice
$\pm$ 0.01.  In this study we adopt $n=5$.

Despite the difference in selection techniques, the stellar mass
distributions of the pairs and post-mergers shown in Figure
\ref{mass_z} are very similar.  However, due to the imposed redshift
cut of the Galaxy Zoo selection, the post-mergers sample is truncated
at significantly lower redshifts than the close pairs.  This is not
\textit{a priori} a problem, since we will describe in the next
section how a control sample is matched in redshift.  The difference
in redshift ranges will only impact our results if there is an underlying
dependence of merger-induced effects on redshift.  We have therefore
repeated all of the analysis presented in this paper and can confirm
that the results are robust to a redshift cut in the pairs at $z=0.1$.
However, since such a redshift cut excludes approximately one-third
of the pairs sample, we have not imposed a $z$ criterion on the pairs, in
order to maximise the statistical sample size.  Similarly, the
distribution of local overdensities differs between the pairs and
post-mergers; this is again tackled by including environment as one of
the parameters in our control sample matching.  Therefore, the differential
changes in, for example, star formation rate, can be compared, even
when the underlying distributions of mass, environment etc. may differ.

\subsection{Control samples}

In previous works we have compiled control samples with a fixed number
(typically $\sim$ 10) of mass-, density- and redshift-matched galaxies
with no close companion.  However, in the current paper, we will be
selecting subsets of galaxies from our main pairs sample (e.g. for
studies of star forming or AGN properties), such that
control samples are more conveniently created dynamically.  The pool
of possible control galaxies (for any subset of pairs) consists of all
galaxies that have no spectroscopic companion within 80 \hkpc\ and
with a relative velocity $\Delta V$ within 10,000 \kms.  Due to the
spectroscopic incompleteness of the SDSS, there will be many
\textit{bona fide} close pairs that are not in our pairs sample, and
are therefore erroneously identified as potential control galaxies.  
In order to avoid using
these unidentified pairs as controls, we utilize the Galaxy Zoo binary
merger sample (Darg et al. 2010), since, unlike our close pairs
sample, Galazy Zoo does not require the companion to have a
spectroscopic redshift.  We therefore additionally reject galaxies
from the control pool if they have Galaxy Zoo merger vote fractions
$f_m > 0$ (see Darg et al. 2010 and Ellison et al. 2013 for more
details).

For a given pair or post-merger galaxy, all of the galaxies from the
control pool within a redshift tolerance $\Delta z$=0.005, a mass
tolerance $\Delta$log M$_{\star}$=0.1 dex and a normalized local density
tolerance $\Delta \delta_5$=0.1 dex are selected as controls.
At least 5 matches are required, a limit which is achieved in $>$ 90 per cent
of cases (the typical number of matches is usually
of the order several hundred).  If less than 5 matches are found, the
tolerance is grown by a further $\Delta z$=0.005 in redshift,
$\Delta$log M$_{\star}$=0.1 dex in stellar mass and  $\Delta \delta_5$=0.1 dex
in normalized local density until the required
number of matches is achieved.  

\subsection{Classification schemes and S/N requirements}

From our complete sample of 10,800 galaxies in pairs and 97
post-mergers, various sub-samples are selected for the analysis
presented in this paper.  These criteria are primarily based on two
parameters.  First, the S/N in the 4 emission lines used in the AGN
diagnostic diagram of Baldwin, Phillips \& Terlevich (1981), namely
[OIII] $\lambda$ 5007, H$\beta$, H$\alpha$ and [NII] $\lambda$ 6584,
hereafter collectively referred to as the BPT lines.  Second, the
choice of AGN or star-forming classification scheme that divides the
BPT diagram.  Ideally, one would adopt a uniform set of S/N and
classification criteria to the whole sample.  In practice, the
different astrophysical tests we will carry out have varying
requirements.  The criteria we adopt in the analysis that follows are
explained and justified in the relevent sections of the paper.
However, it is useful to describe the basic motivations behind our
choices in advance.  

In Sections \ref{sfr_sec} and \ref{agn_sec} we investigate the
star-forming and AGN fractions in our sample.  This experiment
requires only a classification on the BPT diagram.  To select galaxies
that are \textit{dominated} by star-formation we use the criteria of
Kauffmann et al. (2003b)\footnote{According to Stasinska et al.
(2006) star-forming galaxies selected by the Kauffmann et al. (2003b)
criteria may have up to 3 per cent contribution from AGN.}  and to
select galaxies with \textit{some} contribution from AGN we use the
criteria of Stasinska et al.  (2006).  For the classification of
galaxies as star-forming or AGN we require a S/N in all of the BPT
lines of at least 5.  Cid-Fernandes et al. (2010) have made a thorough
investigation of the impact of S/N requirements on the selection of
galaxies within the BPT diagram.  They convincingly demonstrate that
galaxies excluded (or included) by a given S/N criterion are not just
a random sampling of galaxies with poorer quality spectra.  
For galaxies classified as AGN a low S/N more
frequently selects galaxies with Low Ionization Nuclear Emission
Region (LINER-like) spectra.  We therefore require that the S/N of the
BPT lines exceeds 5 for our calculation of AGN and (for consistency)
star-forming fractions.

For the computation of gas phase metallicities, a similarly high S/N
is required (e.g. Kewley \& Ellison 2008), so we follow our previous
work on the metallicity distribution in galaxy pairs and again adopt
S/N$>$5 (Scudder et al. 2012b).  However, since the SFRs themselves
are mostly driven by the strength of the H$\alpha$ line, which is
typically the strongest of the BPT lines by a large margin, we are
able to reduce the collective S/N requirement of the BPT lines to be
S/N$>$1 when quantifying SFRs.  This is similar to the criterion we
have recently adopted in our study of SFRs in wide pairs (Patton et
al. 2013).

Following our previous work, we will trace the impact of the galaxy
interaction relative to the control sample by calculating the offset
($\Delta$), measured on a logarithmic scale, between a given galaxy
metric (such as SFR, which is described more fully in Section
\ref{delta_sfr_sec}) and the median value of its control galaxy
values.  For example,

\begin{equation}
\Delta \rm{SFR} = \log \rm SFR_{\rm pair} - \log \rm SFR_{\rm control}
\end{equation}

such than an offset of $\Delta$ SFR = 1.0 indicates a SFR enhancement
of a factor of 10 above what would be expected for an isolated galaxy of
this mass, environment and redshift.

\section{Triggered Star formation}\label{sfr_sec}

\begin{figure}
\centerline{\rotatebox{0}{\resizebox{9cm}{!}
{\includegraphics{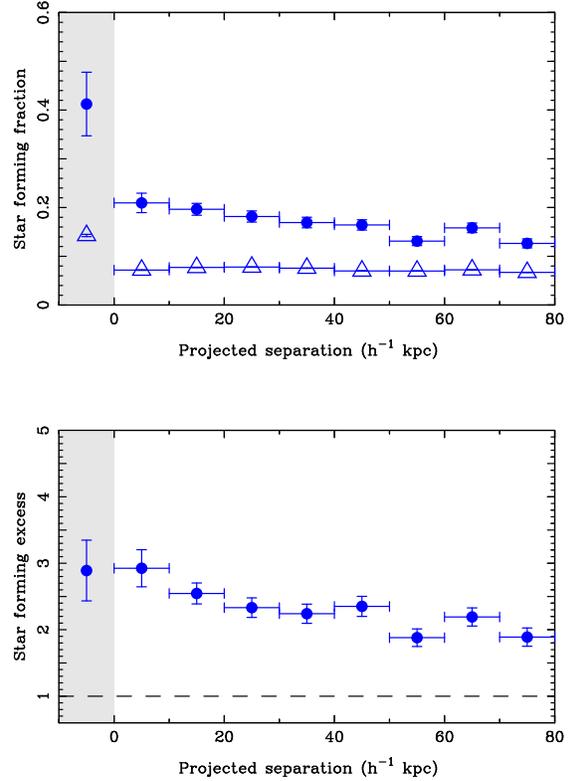}}}}
\caption{\label{sf_frac} The fraction of galaxies classified as
star-forming as a function of projected separation.  In the upper
panel, filled points represent pairs and (in the grey shaded box) post
mergers and open triangles show the star-forming fraction in the
control sample.  Since each galaxy has its own set of controls matched
in redshift, mass and environment, the controls can also be plotted as
a function of \rp\ to test for any underlying dependences.  In the
lower panel we show the ratio of the star-forming fraction in the
pairs and post-mergers relative to their controls. }
\end{figure}

\subsection{The fraction of star-forming galaxies}

We begin our investigation of the properties of post-mergers by
considering the fraction of galaxies that are dominated by
star-formation.  Star-forming galaxies are selected using the criteria
of Kauffmann et al. (2003b) and requiring a S/N $>$ 5 in the BPT lines.
The top panel of Figure \ref{sf_frac} shows the fraction of galaxies
classified as star-forming in both the sample of close pairs and the
post-merger sample.  The open triangles in the upper panel indicate the
fraction of star-forming galaxies in the control samples.  Since each
galaxy has its own set of matched controls, we are able to show the
controls as a function of \rp, so that the control sample will mirror
any changes in the underlying galaxy population.  For example, it can
be seen that the controls that are matched to the post-merger sample
have a significantly higher star-forming fraction than the controls matched
to the pairs.  This is
mainly driven by the typically lower density environments of the
post-merger sample selected by Galazy Zoo compared with our
spectroscopic pairs selected.  The top panel of Figure \ref{sf_frac}
therefore demonstrates the importance of control matching.  However,
for most subsequent metrics of merger-induced effects studied in 
this paper we will show only the values relative to the controls,
akin to the lower panel of  Figure \ref{sf_frac}.

The lower panel of Figure \ref{sf_frac} shows the \textit{excess}
star-forming fraction, i.e. the ratio of the star-forming fractions in
the mergers relative to their controls.  It can be seen that the pairs
exhibit an enhancement in the star-forming fraction out to at least 80
\hkpc, which is the limit of our sample.  This is complementary to the
results of Scudder et al. (2012b) and Patton et al. (2013), who have
recently shown that star formation \textit{rates} remain elevated out
to similar separations (dropping to the control value by \rp\ $\sim$
150 \hkpc).  However, the increase in the star-forming fraction from
\rp\ $\sim$ 80 \hkpc\ to \rp\ $\sim 10$ \hkpc\ is rather modest,
increasing from a factor of $\sim$ 2 to 3 above the control sample over
this separation range.  Similarly, although the raw fraction of star-forming
galaxies in the post-merger sample is higher than in the close
pairs (upper panel of Figure \ref{sf_frac}), so is the fraction of
star-forming galaxies in the matched control sample.  Therefore, the excess of
star-forming galaxies in the post-mergers is actually consistent
with the value in the closest pairs.  This result indicates that
galaxies tend to become classified as star-forming well before final
coalescence.

\subsection{Star formation rate enhancements}\label{delta_sfr_sec}

We next consider the \textit{rates} of star formation within those
galaxies classified as star-forming.  The strength of the H$\alpha$
line relative to the other BPT lines allows us to relax the collective
BPT line S/N constraint to 1, without loss of accuracy in the
determination of the star formation rate (the qualitative trends in
our results are not sensitive to this choice of S/N cut).  The SFRs
are taken from Brinchmann et al. (2004) and are based on template fits
to the SDSS spectra, yielding `fibre' SFRs that are applicable to the
region of the galaxy covered by the 3 arcsecond SDSS aperture.
Brinchmann et al. (2004) also apply a colour-dependent correction for
the light outside of the fibre, in order to determine `total' SFRs.
In this paper, we additionally infer the SFR outside of the 3 arcsec
fibre by subtracting the fibre SFR from the total value.

\begin{figure}
\centerline{\rotatebox{270}{\resizebox{8cm}{!}
{\includegraphics{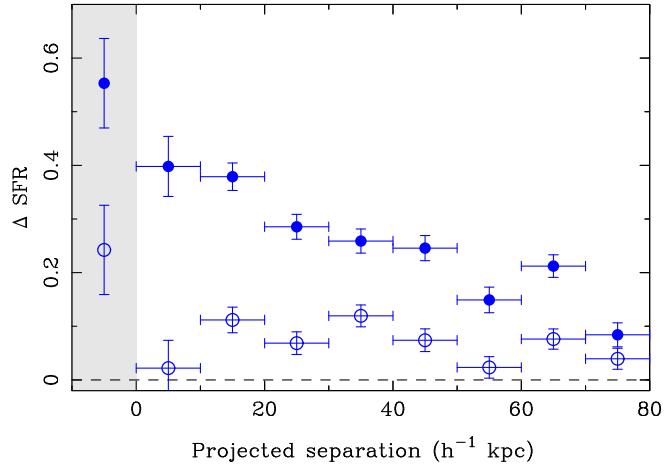}}}}
\caption{\label{dsfr_sf} The enhancement in SFR for close pairs of
galaxies that are classified as star-forming.  Filled points show the
enhancements in the fibre, hollow points show the SFR
enhancements outside of the fibre. The points in the grey shaded box show 
the enhancements
for galaxies in the SDSS post-merger sample presented here. }
\end{figure}

In Figure \ref{dsfr_sf} we show the enhancement in the SFR ($\Delta$
SFR) in both the close pair and post-merger samples relative to their
respective matched control samples.  Both fibre (filled points) and
outer SFRs (open points) show an excess in the close pairs relative to
their controls out to at least 80 \hkpc.  Although previous works have
inferred that the bulk of triggered star formation is centrally
concentrated (Barton et al. 2003; Bergvall et al.  2003; Ellison et
al. 2010; Patton et al. 2011; Schmidt et al. 2013), Figure
\ref{dsfr_sf} shows explicitly that a small amount of star formation
is also triggered in a more spatially extended component.  Star
formation on an extended spatial scale has also been predicted in
simulations (e.g. Powell et al. 2013).  The pre-coalescence SFRs in
our sample of pairs are, on average, a factor of $\sim$ 2 higher than
their control in the fibres and 25 per cent higher in the outer parts
(beyond a few kpc).  The post-mergers show further enhancements in
their SFRs in both the outer and fibre measurements; on average the
post-mergers have SFRs that are higher by 70 per cent (outer) and a
factor of 3.5 (fibre) than their control samples.  These enhancements
exceed (in both cases) the pre-merger values by about 40 per
cent. Therefore, whereas the pre-coalescence phase of the interaction
most strongly affects central star formation, the final merger
increases the SFR more uniformly on a broader spatial scale.

\subsection{The frequency of starbursts}

Whilst Figure \ref{dsfr_sf} measures the evolution of the average
$\Delta$ SFR, it does not tell us about the distribution of 
SFR enhancements,  including how frequently real starburst galaxies are 
formed in the merger process.  Whilst the \textit{average} SFR
enhancement in post-mergers may be only a factor of 3.5, this may be 
due to a mix of galaxies that are already quenched and some that are 
forming stars
very aggressively.  For example, Di Matteo et al. (2007, 2008) show
that galaxies that experience more modest tidal interactions during
their first pericentric passage are able to preserve a higher gas
reservoir for a more intense burst of star formation at coalescence.

In Scudder et al. (2012b) we investigated the cumulative distributions
of $\Delta$ SFR in close pairs as a function of both mass ratio and
separations.  It was shown that whilst close pairs of all mass ratios
and separations may exhibit modest SFR enhancements (up to a factor of
2), enhancements exceeding a factor of 10 were restricted to
approximately equal mass mergers with small projected separations
(although starbursts are rare, even within these criteria).  In this
paper, we take a slightly different approach, and plot the fraction of
star-forming galaxies with enhancements greater than some threshold.
Again, we require that the BPT lines all have a S/N $>$ 1 and the
galaxies are classified as star-forming according to the Kauffmann et
al. (2003b) criteria.  We set the SFR enhancement thresholds to be a
factor of 2, 5 and 10 above the matched control sample (i.e. $\Delta$
SFR $>$ 0.3, 0.7, 1.0). 

 Figure \ref{frac_thresh} shows that the fraction of galaxies
exceeding each of these fibre SFR thesholds increases towards the
smallest separations and peaks in the post-merger sample.
Enhancements of a factor of two are common; 60 -- 70 per cent of the
closest pairs and post-mergers exhibit an increase at this level.  In
the post-mergers, 40 per cent of galaxies have SFR enhancements of at
least a factor of five.  True starbursts are relatively rare, rising
from $\sim$ 5 per cent in the widest separation pairs to $\sim$ 15 per
cent in the closest pairs and 20 per cent in the post-mergers.  These
starburst fractions are in good agreement with the suite of
simulations presented by Di Matteo et al. (2008), who find 13 -- 22
per cent of mergers show a SFR enhancement of at least a factor of 10,
depending on the galaxy inclinations.  The observational data therefore
emphasize the conclusion made in the theoretical work of Di Matteo
et al. (2008), that mergers do not always trigger starbursts.

\subsection{The contribution of coalescence to stellar mass}

\begin{figure}
\centerline{\rotatebox{270}{\resizebox{8cm}{!}
{\includegraphics{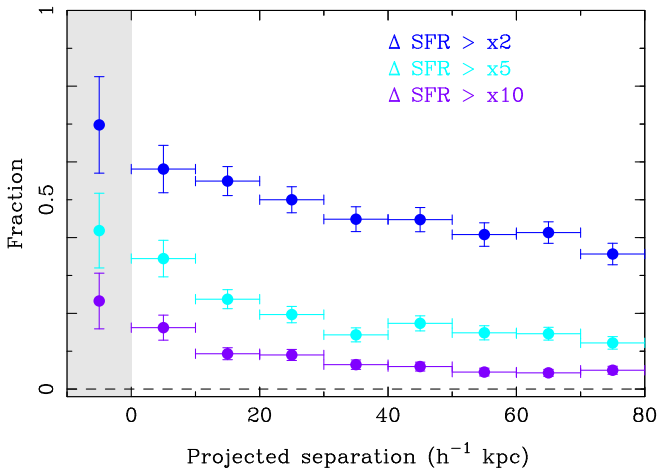}}}}
\caption{\label{frac_thresh} The fraction of galaxies with fibre SFR
enhancements at least a factor of 2, 5 or 10 (blue, cyan and
purple points respectively)
above their matched controls. The points in the grey shaded box show
the enhancements for galaxies in the SDSS post-merger sample. }
\end{figure}

Simulations of galaxy interactions generally conclude that intense
bursts of star formation should occur at, or shortly before, 
coalescence (e.g. Di Matteo
et al. 2007, 2008; Cox et al.  2008; Torrey et al. 2012).  Although
the results in Figures \ref{dsfr_sf} and \ref{frac_thresh} qualitatively
support these predictions, the SFRs at coalescence are predicted to
frequently exceed the pre-merger phase by a factor of 10 or more.
The increase in the average SFR and starburst frequency between the
pre- and post-merger stages shown in Figures   \ref{dsfr_sf} and 
\ref{frac_thresh} are therefore quite modest.  However, it is 
important to take into consideration the timing of the starburst
which is usually both initiated, and reaches its peak, before the 
galaxies fully merge.  These bursts
tend to be intense, but short-lived, such that by a few hundred Myr
after the coalescence the starburst has already largely subsided
(e.g. Fig. 11 of Scudder et al. 2012b).  It is therefore very
unlikely, observationally, to catch a merger during the peak of star
formation activity associated with the final merger.  However, it is
worth noting that since we do detect an enhancement in the SFR in
post-mergers, coalescence has probably occured within this timescale
of a few hundred Myrs, before `quenching' becomes widespread,
a conclusion supported by the metallicities discussed in Section
\ref{metal_sec}.

So far in this section our analysis has focused on metrics associated
with the current star formation activity in pairs and post-mergers.
These metrics are derived from emission lines in HII regions that are
ionized by the presence of young, hot stars.  As argued above, a short
`snapshot' of the current SFR is unlikely catch a short, intense burst of star
formation whose overall impact (in terms of stellar mass produced) may
be equivalent to a long, gentle period of production.  A complementary
demonstration of the recent star formation history, that probes a
more extended timescale, comes from a measurement of the
fraction of stellar mass in a `young' component.  Recently, Mendel et
al. (2013b) have used a spectral decomposition technique to
investigate the properties of quenched galaxies in the SDSS.  The
technique adopts two spectral templates (Bruzual \& Charlot 2003), one
consisting of a 7 Gyr old single stellar population with solar
metallicity, and the other of a 300 Myr old starburst.  The spectral
decomposition determines the fractional contribution to the stellar
mass from the young stellar template, \fy.  

\begin{figure}
\centerline{\rotatebox{0}{\resizebox{9cm}{!}
{\includegraphics{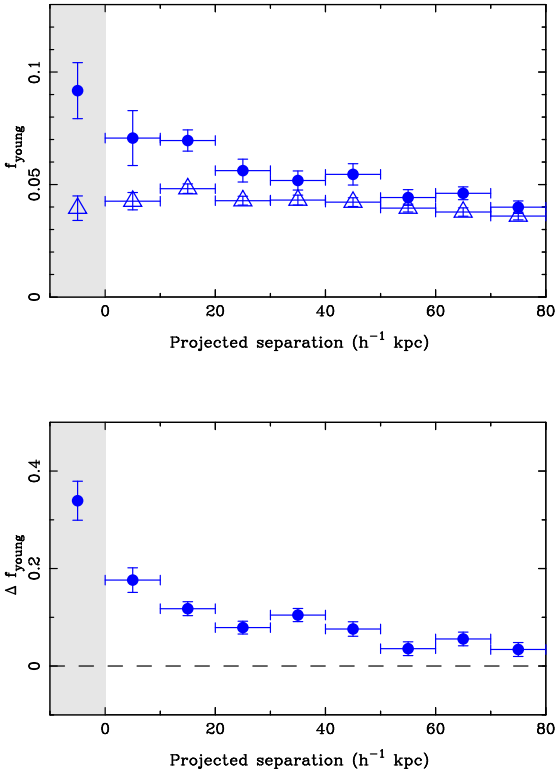}}}}
\caption{\label{dfysf} The fraction of galactic stellar mass
contributed from young stars, \fy, as defined by Mendel et
al. (2013b), for star-forming galaxies.  The upper panel shows the
absolute value of \fy\ in the mergers (filled circles) and control
sample (open triangles).  Since each galaxy has its own set of
controls matched in redshift, mass and environment, the controls can
also be plotted as a function of \rp\ to test for any underlying
dependences. The lower panel shows the enhancement in median \fy\ for
mergers relative to their control sample.  The point in the grey
shaded box shows the enhancement for galaxies in the SDSS post-merger
sample. }
\end{figure}

Figure \ref{dfysf} shows the median difference between \fy\ measured
in the sample of pairs and post-mergers, compared to \fy\ measured in
the matched control samples.  Since \fy\ is not a commonly used
quantity in the literature, in the top panel of Figure \ref{dfysf} we
show the absolute values of this quantity for reference.  We include
all of the galaxies for which a spectral decomposition is successful
and are classified as star-forming (Kauffmann et al. 2003b).  The
enhancement in \fy\ can be used to deduce the increase in the stellar
mass due to star formation over $\sim$ 300 -- 500 Myr timescales.
This is much shorter than the full interaction timescale probed by the
close pairs, so changes in \fy\ in pairs does not provide a complete
inventory of new stellar mass in the pre-merger phase.  However, as we
argue below, the time since coalescence of our post-merger sample is
likely to be sufficiently short that $\Delta$ \fy\ for this population
will encapsulate the integrated stellar mass formed during
this final stage.  Figure \ref{dfysf} shows a median enhancement
of \fy\ in post-mergers that corresponds to an approximate doubling of
the fraction of mass in young stars, relative to a matched control
sample.  Therefore, by using \fy\ as a complementary measure of recent
(but not instantaneous) star formation, we can see that coalescence
plays an important role in stellar mass production, despite the short
timescale that these bursts are predicted to endure.

\medskip

In summary of this section, we have found that star formation is
triggered early on in the interaction process.  Whilst coalescence
does induce further star formation, the SFRs are (on average) only
modestly higher than during the pre-merger phase of the interaction.
However, instantaneous SFRs do not provide the complete picture of
star formation throughout the merger sequence.  Indeed the combination
of prolonged SFR enhancements out to wide separations (Patton et
al. 2013) and the results presented here, indicate that typical
post-pericentric bursts of star formation may endure for at least a
Gyr, but that coalescence induced star formation may be more
short-lived.  Nonetheless, the short, intense burst of star formation
that typifies coalescence can approximately double the fraction of
mass in young stars.

\section{Changes in metallicity}\label{metal_sec}

\begin{figure}
\centerline{\rotatebox{270}{\resizebox{8cm}{!}
{\includegraphics{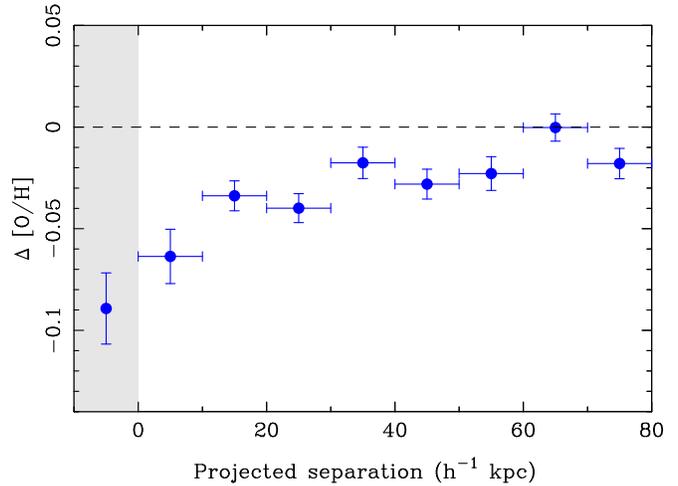}}}}
\caption{\label{doh} The change in metallicity for close pairs of galaxies
relative to their control sample.  The point in the
grey shaded box shows the metallicity offset for galaxies in the SDSS 
post-merger sample. }
\end{figure}

Following Scudder et al. (2012a,b), we calculate the metallicity of
star-forming galaxies according to the formalism of Kewley \& Dopita
(2002), as recently re-assessed by Kewley \& Ellison (2008).  In order
to be consistent with our previous work, we require that galaxies are
classified as star-forming according to the criteria of Kauffmann et
al. (2003b) and have a S/N $>$ 5 in H$\alpha$, H$\beta$, [OII]
$\lambda$ 3727, [OIII] $\lambda\lambda$ 4959, 5007 and [NII] $\lambda$ 6584.
In Figure \ref{doh} the change in metallicity, relative to
the control sample, is plotted for both the close pairs and 
post-mergers. Figure \ref{doh} shows the
now familiar trend in the pairs of a projected separation
dependence that remains offset from the control sample out to at
least 80 \hkpc.  The close pairs are typically 0.03 dex more metal-poor
than the control sample, with the largest offset at the closest
separations ($\Delta$ [O/H] $ \sim -0.07$).  The post-mergers are
even more metal-poor, with a median $\Delta$ [O/H] $ \sim -0.09$.
Since metal-enrichment will follow a starburst, simulations predict
that the metallicity dilution is most extreme within $\sim$ 100 Myr
of final coalescence, before it recovers to (or exceeds) its pre-merger
value (e.g. Montuori et al. 2010; Scudder et al. 2012b; Torrey et al.
2012).  The low observed metallicities in our post-merger sample
support the conclusion drawn from the star formation rates, that
coalescence is likely to have occurred in these galaxies relatively 
recently.

\section{AGN in post-mergers}\label{agn_sec}

\subsection{The fraction of AGN}

In the same way that we calculated the fraction of galaxies classified
as star-forming in Figure \ref{sf_frac}, we will now compute the
fraction of galaxies that exhibit some contribution from AGN in their
emission spectrum.  The result of this calculation is particularly
sensitive to the choice of BPT line S/N, as lowering the detection
threshold includes more LINER-like objects (e.g. Cid-Fernandes et
al. 2010; Ellison et al. 2011).  Due to the uncertain nature of LINERs
(e.g. Stasinska et al.  2008; Cid Fernandes et al. 2011; Yan \&
Blanton 2012), it is preferable to restrict ourselves to Seyfert-like
spectra and we therefore impose a S/N $>$ 5 requirement on the BPT
lines for this basic classification.  We adopt the Stasinska et al.
(2006) classification in order to be consistent with our previous work
(Ellison et al. 2011).  However, we stress that the Stasinska et
al. (2006) criteria select galaxies with a wide range of AGN
contribution and include composite objects that also have significant
star formation.  Nonetheless, the Stasinska et al. (2006) selection is
appropriate for determining the fraction of galaxies with measurable
AGN contribution and it was shown by Ellison et al. (2011) that the
basic trend of increasing AGN fraction with decreasing projected
separation is not dependent on the choice of diagnostic.

\begin{figure}
\centerline{\rotatebox{270}{\resizebox{8cm}{!}
{\includegraphics{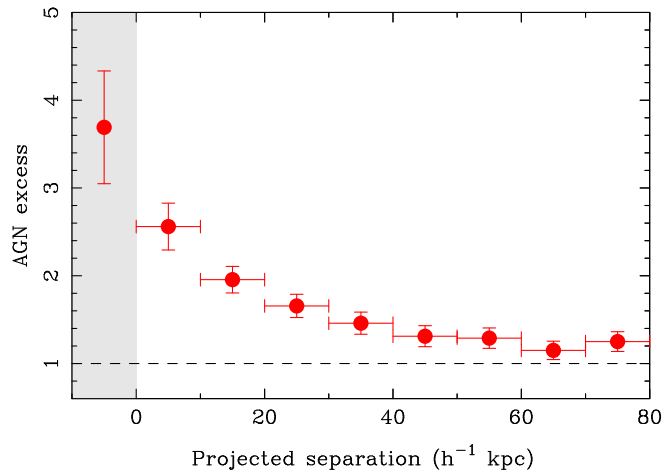}}}}
\caption{\label{agn_frac} The AGN excess, defined as the fraction of
SDSS galaxies that are classified as an AGN according to the
definition of Stasinska et al. (2006), relative to the fraction of AGN
of their respective controls. The point in the grey shaded box shows
the AGN excess for galaxies in the SDSS post-merger sample.}
\end{figure}

In Figure \ref{agn_frac} we plot the AGN excess, defined to be the
fraction of galaxies with an AGN contribution in both the close pairs
and post-merger samples, relative to the AGN fraction in their
respective controls.  The AGN excess is therefore a measure of how
much more often a galaxy in a merger exhibits the signature of an AGN
compared to a non-merger of the same stellar mass, redshift and
environment.  The qualitative picture for the close pairs is very
similar to that presented in Ellison et al.  (2011); a gradual rise in
the frequency of AGN, which peaks at $\sim$ 2.5 times the control
value at the smallest separations. However, whereas in Ellison et
al. (2011) we found an AGN excess only out to \rp\ $\sim$ 40 \hkpc, in
the present analysis we find a statistical excess out to at least 80
\hkpc, where AGN are more common by 10--20 per cent than in the
control sample.  There are a few differences between our current work
and that of Ellison et al. (2011) that may explain the AGN excess that
is seen out to large separations.  The first is the different range of
mass ratios\footnote{There is also a slightly different \dv\ cut, 300
\kms\ used here, compared with 200 \kms\ in Ellison et al. (2011).}
used in this study (1:4) and Ellison et al. (2011, 1:10).  Second,
Ellison et al. (2011) had a much smaller control sample than this
work, only 10 control galaxies per pair, compared to several hundreds
used here.  However, having experimented with the two samples, neither
of these aspects of the analysis seems to be the cause for the wide
separation AGN excess seen in Figure \ref{agn_frac}.  The main cause
appears to be the control matching parameters.  Ellison et al.  (2011)
matched only in mass and redshift, whereas this paper also matches in
environment.  Therefore, as shown in the mock surveys produced by
Perez et al. (2009), whilst mass is arguably the most important
parameter for matching within a control sample, environmental matching
provides a small, but measurable, improvement in sensitivity.

The main contribution of this work is to add the post-merger galaxies
to Figure \ref{agn_frac}.  In contrast to the fraction of star-forming
galaxies shown in Figure \ref{sf_frac}, there is a clear increase in
the AGN fraction post-coalescence.  The post-mergers have an AGN
frequency that exceeds their matched control sample by a factor
of 3.75.  This confirms theoretical expectations that the accretion rate
is increased the most during the final merger (Di Matteo et al. 2005;
Springel et al. 2005; Johansson et al. 2009).  A similar enhancement
in AGN fraction was found in the spheroidal post-merger sample
of Carpineti et al. (2012), but that work was not able to trace
the increase in AGN occurence throughout the merger sequence.
The high post-merger AGN fraction in our sample is also consistent 
with the observation that LIRGs and
ULIRGs frequently exhibit nuclear activity (e.g.  Yuan, Kewley \&
Sanders 2010; Petric et al. 2011; Ellison et al. 2013).  
One implication of the enhanced
incidence of AGN shown in Figure \ref{agn_frac} is that attempts to
quantify the merger-induced AGN fraction from the identification of
close pairs will likely underestimate the true value.  Although AGN
\textit{may} be triggered during the early phase of an interaction
(see also Rogers et al. 2009; Ellison et al. 2011; Silverman et
al. 2011; Koss et al. 2012; Smith et al. 2012; Liu et al. 2012;
Sabater et al. 2013), the AGN rate peaks post-coalescence. A
similar conclusion has recently been reached by Canalizo \& Stockton
(2013) for a sample of low redshift ($z \sim 0.2$) quasars, which
exhibit the signature of an intermediate-age starburst, in addition
to the on-going AGN activity.

\subsection{Black hole accretion rate}

In order to quantify the black hole
accretion rate in SDSS AGN, numerous studies
have used the luminosity of the [OIII] emission line (e.g.Kauffmann et
al. 2003b; Heckman et al. 2004; Chen et al. 2009; Liu et al. 2012).
However, since star formation can also contribute to the [OIII] line
flux, some studies have attempted to remove this possible
contamination (e.g.  Kauffmann \& Heckman 2009; Wild, Heckman \&
Charlot 2010).  In this paper, we have taken the approach of selecting
galaxies which should have minimal (relative) contribution from star
formation by selecting for this portion of the analysis galaxies that
are classified as AGN by Kewley et al. (2001).  These galaxies are
above the `maximum starburst' line computed from grids of models with
varying metallicities and ionization parameters.  Kauffmann \& Heckman
(2009) estimate that only 10 -- 20 per cent of the [OIII] flux should
be contributed from star formation in AGN classified by Kewley et
al. (2001).  More recent tests by Wild et al. (2010) imply AGN
contributions of close to unity, indicating that although the Kewley
et al. (2001) was not necessarily designed as a `pure AGN' line
(e.g. see the discussion in Stasinska et al. 2006), it is effective at
isolating galaxies with minimal contamination of the [OIII] line by
star formation.  However, the stringent nature of this AGN
classification greatly reduces the size of the AGN sample and
increases the statistical uncertainty of our results.  Lowering the
required S/N can partly offset this problem, but this comes at the
expense of LINER contamination, as described in previous sections.  
The solution we
adopt is to indeed lower the BPT S/N requirement to one, but also
require detections of [OI] $\lambda$ 6300 and [SII] $\lambda$ 6717, 6731.
We can then apply the LINER diagnostic of Kewley et al. (2006),
allowing us to construct a sample of Seyfert-dominated spectra.

\begin{figure}
\centerline{\rotatebox{270}{\resizebox{8cm}{!}
{\includegraphics{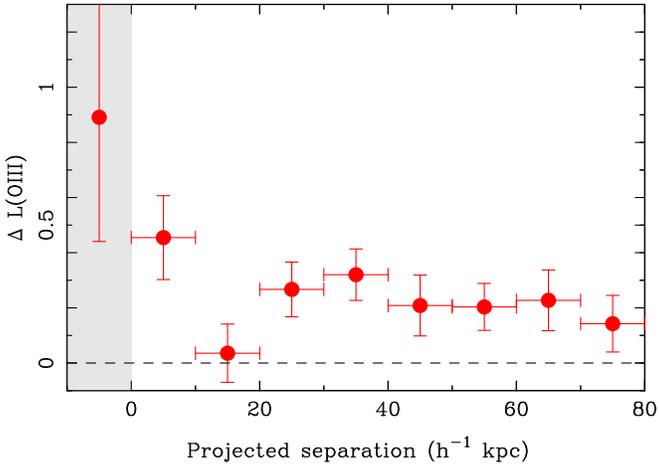}}}}
\caption{\label{do3} The enhancement in the [OIII] luminosity for
close pairs of AGN galaxies relative to their control sample.  The point
in the grey shaded box shows the enhancement for galaxies in the SDSS
post-merger sample. AGN are classified according to the criterion
of Kewley et al. (2001) in order to minimize contamination from star-formation.
LINERs are excluded using the criteria of Kewley et al. (2006).}
\end{figure}

Figure \ref{do3} shows the enhancement in the [OIII] luminosity ($\Delta$
L[OIII]) for both the close pairs and post-merger galaxies.  The
statistics are noisier than for the other metrics investigated in
this paper due to the small number of `pure' Seyferts in our
sample.  Nonetheless, there is a clear enhancement in the  [OIII] 
luminosity by a factor of $\sim$ 2 -- 3 for close pairs separated
by less than 50 \hkpc, consistent with, for example, the recent work 
by Liu et al. (2012).  The post-mergers show the largest accretion
rate enhancement, albeit with a large error bar, as predicted by
simulations of major mergers (e.g. Johansson et al. 2009).

\medskip

\section{Discussion}

Recently, Treister et al. (2012) found a correlation between the
fraction of AGN associated with mergers and bolometric luminosity,
claiming that mergers only trigger the most luminous AGN.  It is
therefore of note that despite the enhancement in the AGN fraction and
the [OIII] luminosity in our pairs and post-mergers
(Figs. \ref{agn_frac} and \ref{do3}), the absolute luminosities are
still fairly modest.  The correction of the [OIII] luminosity to a
bolometric luminosity is highly uncertain, and depends on (among other
factors) how (or if) dust is accounted for and whether the AGN is a
type I or type II (e.g. Heckman et al. 2005).  If we adopt L$_{\rm
bol}$ = 600 $\times$ L([OIII]), as appropriate for dust-corrected type
II AGN (Kauffmann \& Heckman 2009), we find typical bolometric
luminosities 43 $<$ log L$_{\rm bol} < 45$ erg/s, with very few above
this upper bound.  The AGN in our merger sample are therefore not high
luminosity objects.  Whilst it is true that the highest luminosity AGN
appear to be dominated by mergers (e.g. Ramos-Almeida et al. 2011;
2012), interactions can clearly result in AGN that can be observed to
have relatively low luminosities.

The association between mergers and AGN of different luminosities is
complicated by the expectation of highly stochastic accretion rates.
It has been shown in several recent works that AGN feeding by a
variety of mechanisms may proceed at very low rates for extended
periods of time, interspersed with short-lived high accretion rate
episodes (Novak, Ostriker \& Ciotti 2011; Gabor \& Bournaud 2013;
Stickley \& Canalizo 2013).  The enhanced frequency of AGN observed in
our sample of pairs indicates that merger-induced AGN triggering can
occur well before coalescence (see also Ramos-Almeida et al. 2011),
although these AGN typically have modest luminosities.  This is in
good agreement with the simulations of Stickley \& Canalizo (2013) who
predict that small amounts of gas are already reaching the nucleus as
galaxies recede from their first pericentric passage.  In that model,
the highest accretion rates are achieved after coalescence, so we
might expect that the sample of post-mergers would include some of the
highest accretion rate objects.  

Although the AGN frequency peaks in the post-merger sample presented
here, the bolometric luminosities of post-merger AGN are still modest
with values log L$_{\rm bol} < 45$ erg/s.  There are several possible
reasons for the lack of high luminosity AGN in post-merger sample.
First, despite selection from one of the largest galaxy surveys
currently available, the post merger sample is still fairly small, and high
luminosity AGN are expected to be both intrinsically rare and
short-lived.  Not all galaxies may have the requisite gas supply to
feed the central supermassive black holes at high fractions of the
Eddington rate.  Finally, we have argued that the post-mergers in
our sample are likely to have only recently (within a few hundred
Myrs) coalesced, so there may not yet have been sufficient time
for the bulk of gas to reach the nucleus.

The results presented in Section \ref{agn_sec}, which include both the
frequency of AGN and the black hole accretion rate, show both
similarities and differences with the conclusions concerning triggered
star formation in Section \ref{sfr_sec}.  On the one hand, both AGN
and star-forming fractions (and the associated intensity of these
processes as measured by accretion and star formation rates) show
enhancements out to wide separations.  The simulations presented in
Scudder et al. (2012b) and Patton et al.  (2013) indicate that such
wide separation enhancements are due to gas flows triggered at
pericentric passage that are seen over the course of the merger's
dynamical timescale.  However, in contrast to the frequency of star
forming galaxies, and their SFRs, which increase only modestly
post-merger, the AGN frequency and accretion rates rise significantly
after coalescence.  It seems that whilst both star formation and black
hole accretion are enhanced throughout the merger sequence, the former
process responds strongly to the interaction already in the early
stages, whereas the AGN phase reaches its peak after
coalescence. Although simulations have long predicted this general
scenario, including the delay between the peak in these two processes
(Hopkins 2012), the homogeneous analysis of the large pre- and
post-merger samples presented here, has provided the first complete
observational picture of mergers.

\section{Conclusions}

We have presented a sample of 97 post-merger galaxies selected from
the Galaxy Zoo DR7 catalogue of Darg et al. (2010) which we have
combined with our existing sample of 10,800 galaxies in spectroscopic
close pairs (\rp\ $\le$ 80 \hkpc, \dv\ $\le 300$ \kms).  Control
samples have been constructed by matching in mass, redshift and local
environment; several hundred control galaxies are typically identified
for every pair or post-merger galaxy.  Analysed in a homogeneous way,
this combined sample allows us to trace the effects of the interaction
throughout the merger sequence and quantify the relative importance of
coalescence.  Our main conclusions are as follows.

\begin{enumerate}

\item  The fraction of star-forming galaxies is enhanced, 
relative to the control sample, at a similar level for close pairs 
and post-mergers (Figure \ref{sf_frac}).  
Whereas the early interaction phase can apparently turn non-star-forming
into star-forming (by our definition) galaxies, the final coalescence
has a minimal additional effect.  That is, most star-forming post-mergers were
likely to already have been star-forming pre-coalescence.

\item  In the pre-merger phase, we confirm previous results that
the central SFR is enhanced by a factor of a few, on average. We
also show that the SFR is enhanced outside of the central few kpc
by 25 per cent (Figure \ref{dsfr_sf}).  Post-coalescence, both
the inner and outer SFRs are enhanced by a further 40 per cent,
leading to a fibre star formation rate enhancement in the post-mergers
of a factor of 3.5.

\item The fraction of starburst galaxies, with SFR enhancements
at least a factor of 10 larger than their controls, peaks in the
post-mergers at approximately 20 per cent (Figure \ref{frac_thresh}). 

\item  The star formation that is triggered at coalescence
approximately doubles the rate of stellar mass growth, relative to
a sample of matched control galaxies (Figure \ref{dfysf}).  

\item  The metallicities of close pairs are lower than in the control
sample out to the widest separations in our sample, with a smooth
correlation between metal-deficiency and \rp.  The post-mergers
are the most metal poor, exhibiting a median $\Delta$ [O/H] $= -0.09$,
see Figure \ref{doh}.

\item  The persistence and relative elevation of enhanced SFRs and 
metal deficiencies indicates that the galaxies in our post-merger
sample are likely to have coalesced within the last few hundred Myr.

\item  In contrast to the star-forming fractions, the fraction of
AGN in post-mergers is significantly higher than in the close pairs,
and higher than the control by a factor of 3.75
(Figure \ref{agn_frac}).  Therefore, although AGN may be triggered
pre-coalescence, the final merger has the highest impact on black hole 
accretion.  This is demonstrated directly in Figure \ref{do3}, where
it is shown that the enhancement in the [OIII] luminosity is
greatest in the post-mergers.

\end{enumerate}

\section*{Acknowledgments} 

SLE and DRP acknowledge the receipt of NSERC Discovery grants which
funded this research. Thanks to Asa Bluck, Jorge Moreno and Paul Torrey for
many enjoyable conversations on galaxy mergers and suggestions
for this work.  We are grateful to Daniel Darg for providing us
with his Galaxy Zoo post-merger sample, which formed the basis of
this work.

Funding for the SDSS and SDSS-II has been provided by the Alfred
P. Sloan Foundation, the Participating Institutions, the National
Science Foundation, the U.S. Department of Energy, the National
Aeronautics and Space Administration, the Japanese Monbukagakusho, the
Max Planck Society, and the Higher Education Funding Council for
England. The SDSS Web Site is http://www.sdss.org/.

The SDSS is managed by the Astrophysical Research Consortium for the
Participating Institutions. The Participating Institutions are the
American Museum of Natural History, Astrophysical Institute Potsdam,
University of Basel, University of Cambridge, Case Western Reserve
University, University of Chicago, Drexel University, Fermilab, the
Institute for Advanced Study, the Japan Participation Group, Johns
Hopkins University, the Joint Institute for Nuclear Astrophysics, the
Kavli Institute for Particle Astrophysics and Cosmology, the Korean
Scientist Group, the Chinese Academy of Sciences (LAMOST), Los Alamos
National Laboratory, the Max-Planck-Institute for Astronomy (MPIA),
the Max-Planck-Institute for Astrophysics (MPA), New Mexico State
University, Ohio State University, University of Pittsburgh,
University of Portsmouth, Princeton University, the United States
Naval Observatory, and the University of Washington.


\begin{thebibliography}{}
\small
\itemindent -0.48cm


\bibitem[Alonso et al. (2006)]{alo06}
        Alonso, M. S.,  Lambas, D. G., Tissera, P. B., Coldwell, G.,
	 2006, MNRAS, 367, 1029

\bibitem[Alonso et al. (2007)]{alo07}
        Alonso, M. S.,  Lambas, D. G., Tissera, P. B., Coldwell, G.,
	  2007, MNRAS, 375, 1017

\bibitem[Alonso et al. (2012)]{alo12}
        Alonso, M. S., Mesa, V. Padilla, N. Lambas, D. G., 
	2012, A\&A, 539, 46

\bibitem[Alonso et al. (2010)]{alo10}
        Alonso, M. S., Michel-Dansac, L., Lambas, D. G., 	
	2010, A\&A, 514, 57

\bibitem[Baldwin, Phillips \& Terlevich (1981)]{bpt81}
        Baldwin, J. A,; Phillips, M. M., Terlevich, R., 1981, PASP, 93, 5

\bibitem[Barnes \& Hernquist (1996)]{bh96}
        Barnes, J. E., \& Hernquist, L., 1996, ApJ, 471, 115

\bibitem[Barton, Geller \& Kenyon (2000)]{bgk00}
        Barton, E. J., Geller, M. J., \& Kenyon, S. J., 2000, ApJ, 530, 660

\bibitem[Barton, Geller \& Kenyon (2003)]{bgk03} %colour central colours
        Barton, E. J., Geller, M. J., \& Kenyon, S. J., 2003, ApJ, 582, 668

\bibitem[Bergvall et al. (2003)]{berg03}
        Bergvall, N., Laurikainen, E., Aalto, S., 2003, A\&A, 405, 31

\bibitem[Brinchmann et al. 2004]{bri04} % sdss sfr
        Brinchmann, J., Charlot, S., White, S. D. M., Tremonti, C., 
	Kauffmann, G., Heckman, T., Brinkmann, J.,2004, MNRAS, 351, 1151  

\bibitem[Bruzual \& Charlot (2003)]{bc03}
        Bruzual, G., \& Charlot, S., 2003, MNRAS, 344, 1000	

\bibitem[Canalizo \& Stockton (2013)]{cs13}
         Canalizo, G., \& Stockton, A.,  2013, ApJ, 772, 132

\bibitem[Carpineti et al (2012)]{car12}
         Carpineti, A., Kaviraj, S., Darg, D., Lintott, C., Schawinski, 
	 K., Shabala, S.,  2012, MNRAS, 420, 2139

\bibitem[Casteels et al. (2013)]{cas13}
        Casteels, K. R. V., Bamford, S. P., Skibba, R. A., Masters, K. L., 
        Lintott, C. J., Keel, W. C., Schawinski, K., Nichol, R. C., 
	Smith, A. M., 2013, MNRAS, 429, 1051

\bibitem[Chen et al. (2009)]{chen09}
        Chen, Y.-M., Wang, J.-M., Yan, C.-S., Hu, C., Zhang, S.,
	2009, ApJ, 695, L130

\bibitem[Chung et al. (2013)]{ch13}
         Chung, J., Rey, S.-C., Sung, E.-C., Yeom, B.-S.,; Humphrey, A., 
	 Yi, W., Kyeong, J., 2013, ApJ, 767, L15

\bibitem[Cid-Fernandes et al. (2010)]{cf10}
          Cid Fernandes, R., Stasinska, G., Schlickmann, M. S., 
	  Mateus, A., Vale Asari, N., Schoenell, W., Sodre, L.,
	   2010, MNRAS, 403, 1036

\bibitem[Cid-Fernandes et al. (2011)]{cf11}
          Cid Fernandes, R., Stasinska, G., Mateus, A.,
	   Vale Asari, N., 2011, MNRAS, 413, 1687

\bibitem[Cotini et al. (2013)]{cot13}
        Cotini, S., et al., 2013, MNRAS, 431, 2661

\bibitem[Cox et al. (2006)]{cox06}
         Cox, T. J., Jonsson, P., Primack, J. R., Somerville, R. S.,
	 2006, MNRAS, 373, 1013

\bibitem[Cox et al. (2008)]{cox08}
        Cox, T. J., Jonsson, P., Somerville, R. S., Primack, J. R.,
	Dekel, A., MNRAS, 2008, 384, 386

\bibitem[Darg et al. (2010a)]{darg10a} % merger props
        Darg, D. W., et al., 2010, MNRAS, 401, 1552

\bibitem[Debuhr, Quataert \& Ma (2011)]{dqm11}
        Debuhr, J., Quataert, E., \& Ma, C.-P., 
	2011, MNRAS, 412, 1341

\bibitem[Di Matteo et al. (2008)]{dim08}
        Di Matteo, P., Bournaud, F., Martig, M., Combes, F., Melchior, 
	A.-L., Semelin, B., 2008, A\&A, 492, 31

\bibitem[Di Matteo et al. (2007)]{dim07}
        Di Matteo, P., Combes, F., Melchior, A.-L., Semelin, B.,
	2007, A\&A, 468, 61

\bibitem[Di Matteo et al. (2005)]{dim05}
        Di Matteo, T., Springel, V., Hernquist, L., 2005, Nature, 433, 604

\bibitem[Domingue et al. (2009)]{dom09}
        Domingue, D. L., Xu, C. K., Jarrett, T. H., Cheng, Y
	2009, ApJ, 695,1559

\bibitem[Ellison et al. (2011)]{agn}
         Ellison, S. L., Patton, D. R., Mendel, J. T.,  Scudder, 
	 J. M., 2011, MNRAS, 418, 2043

\bibitem[Ellison et al. (2008)]{sle08a}
         Ellison, S. L., Patton, D. R., Simard, L., McConnachie, A. W.,
	 2008 AJ, 135, 1877

\bibitem[Ellison et al. (2010)]{sle10}
         Ellison, S. L., Patton, D. R., Simard, L., McConnachie, A. W.,
	 Baldry, I. K., Mendel, J. T.,
	 2010, MNRAS 407, 1514. 

\bibitem[Ellison et al. (2013)]{sle13}
         Ellison, S. L.,  Mendel, J. T., Scudder, J. M., Patton, D. R., 
	 Palmer, M. J. D., 2013, MNRAS, 430, 3128

\bibitem[Gabor \& Bournaud (2013)]{gb13}
        Gabor, J. M., \& Bournaud F., 2013, MNRAS, 434, 606G

\bibitem[Heckman et al. (2004)]{heck04}
        Heckman, T. M., Kauffmann, G., Brinchmann, J., Charlot, S., 
	Tremonti, C., White, S. D. M.,  2004, ApJ, 613, 109

\bibitem[Heckman et al. (2005)]{heck05}
         Heckman, T. M., Ptak, A., Hornschemeier, A., Kauffmann, G.,
	 2005, ApJ, 634, 161

\bibitem[Hopkins (2012)]{hop12}
        Hopkins, P., 2012, MNRAS, 420, L8

\bibitem[Hwang et al. (2010)]{hwa10}
        Hwang, H. S., Elbaz, D., Lee, J. C., Jeong, W.-S., Park, C., 
	Lee, M. G., Lee, H. M., 2010, A\&A, 522, 33

\bibitem[Hwang et al. (2011)]{hwa11}
        Hwang, H. S.,  et al. 2011, A\&A, 535, 60

\bibitem[Jogee et al. (2009)]{jog09}
        Jogee, S., et al., 2009, ApJ, 697, 1971

\bibitem[Johansson et al. (2009)]{jnb09}
        Johansson, P. H., Naab, T., Burkert, A., 2009, ApJ, 690, 802

\bibitem[Kartaltepe et al. (2010b)]{kar10b} % science
        Kartaltepe J. S., et al., 2010, ApJ, 721, 98

\bibitem[Kauffmann et al. (2003)]{kau03a} %sdss masses
	Kauffmann, G., et al., 2003a, MNRAS, 341, 33

\bibitem[Kauffmann et al. (2003)]{kau03b} %agn
	Kauffmann, G., et al., 2003b, MNRAS, 346, 1055

\bibitem[Kauffmann \& Heckman (2009)]{kh09}
        Kauffmann, G., \& Heckman, T. M., 2009, MNRAS, 397, 135

\bibitem[Kennicutt et al. (1987)]{ken87}
        Kennicutt, R. C., Jr., Roettiger, K. A., Keel, W. C., van der Hulst, 
	J. M., Hummel, E., 1987, AJ, 93, 1011

\bibitem[Kewley \& Dopita (2002)]{kd02}
        Kewley, L. J., \& Dopita, M. A., 2002, ApJS, 142, 35

\bibitem[Kewley \& Ellison (2008)]{ke08}
        Kewley, L. J., \& Ellison, S. L., 2008, ApJ, 681, 1183

\bibitem[Kewley et al. (2006)]{kewl06}
         Kewley, L. J., Geller, M. J., Barton, E. J.,
	  2006, AJ, 131, 2004

\bibitem[Kewley et al. (2006)]{kew06}
         Kewley, L. J., Groves, B., Kauffmann, G., Heckman, T.,
	 2006, MNRAS, 372, 961

\bibitem[Knapen \& James (2009)]{kj09}
        Knapen, J. H., \& James, P. A., 2009, ApJ, 698, 1437

\bibitem[Koss et al. (2012)]{koss2012}
        Koss, M., Mushotzky, R., Treister, E., Veilleux, S., Vasudevan, 
	R., Trippe, M., 2012, ApJ, 746, L22

\bibitem[Koss et al. (2010)]{koss2010}
        Koss, M., Mushotzky, R., Veilleux, S., Winter, L.,
	 2010, ApJ, 716, L125

\bibitem[Lambas et al. (2012)]{lam12} %major/minor
        Lambas, D. G., Alonso, S., Mesa, V., O'Mill, A. L.,
	2012, A\&A, 539, 45

\bibitem[Lanz et al. (2013)]{lanz13}
        Lanz, L., et al., 2013, ApJ, 768, 90

\bibitem[Larson \& Tinsley (1978)]{lt78}
        Larson, R. B., \& Tinsley, B. M.,  1978, ApJ, 219, 46

\bibitem[Li et al. (2008a)]{li08a} %sf
        Li, C., Kauffmann, G., Heckman, T. M., Jing, Y. P., White, S. D. M.,
	2008a, MNRAS, 385, 1903

\bibitem[Li et al. (2008b)]{li08b} %agn
        Li, C., Kauffmann, G., Heckman, T. M., White, S. D. M.,
	Jing, Y. P., 2008b, MNRAS, 385, 1915

\bibitem[Lintott et al. (2008)]{gz}
        Lintott, C. J., et al., 2008, MNRAS, 389, 1179

\bibitem[Liu et al. 2012]{liu12} % AGN pair properties
        Liu, X., Shen, Y., Strauss, M. A.,
	2012, ApJ, 745, 94

\bibitem[Lotz et al. (2008)]{lotz08}
        Lotz, J. M., Jonsson, P., Cox, T. J., Primack, J. R.,
	 2008, MNRAS, 391, 1137

\bibitem[Lotz et al. (2010)]{lotz10} % martio observability
        Lotz, J. M., Jonsson, P., Cox, T. J., Primack, J. R.,
	2010, MNRAS, 404, 575

\bibitem[Mendel et al. (2013)]{mass}
        Mendel, J. T., Palmer, M. J. D., Simard, L., Ellison, S. L.,  
	Patton, D. R., 2013a, ApJS, submitted 

\bibitem[Mendel et al. (2013)]{quench}
        Mendel, J. T.,  Simard, L., Ellison, S. L.,  
	Patton, D. R.,  2013b, MNRAS, 429, 2212

\bibitem[Michel-Dansac et al. (2008)]{md08}
        Michel-Dansac, L., Lambas, D. G., Alonso, M. S., Tissera, P.,
	2008, MNRAS, 386, 82

\bibitem[Mihos \& Hernquist (1994)]{mh94}
        Mihos, C., \& Hernquist, L., 1994, ApJ, 425, L13

\bibitem[Mihos \& Hernquist (1996)]{mh96}
        Mihos, C., \& Hernquist, L., 1996, ApJ, 464, 641

\bibitem[Montuori et al. (2010)]{mon10}
         Montuori, M., Di Matteo, P., Lehnert, M. D., Combes, F., 
	 Semelin, B.,  2010, A\&A, 518, 56

\bibitem[Nikolic, Cullen \& Alexander (2004)]{nca04}
        Nikolic, B., Cullen, H., Alexander, P., 2004, MNRAS, 355, 874

\bibitem[Novak, Ostriker \& Ciotti (2011)]{noc11}
        Novak, G. S., Ostriker, J. P., \& Ciotti, L., 2011, ApJ, 737, 26

\bibitem[Park \& Choi (2009)]{pc09}
        Park, C., \& Choi, Y.-Y., 2009, ApJ, 691, 1828

\bibitem[Patton \& Atfield (2008)]{pa08}
         Patton, D. R., \& Atfield, J. E., 2008, ApJ, 685, 235

\bibitem[Patton et al. (2011)]{dave11}
         Patton, D. R., Ellison, S. L.,  Simard, L., McConnachie, A. W.,
	 Mendel, J. T.,
	 2011, MNRAS, 412, 591 
	
\bibitem[Patton et al. (2013)]{letter}
        Patton, D. R., Torrey, P., Ellison, S. L., Mendel, J. T.,
        Scudder, J. M., 2013, MNRAS, 433, L59

\bibitem[Peeples et al. (2009)]{peep09} % metal poor
        Peeples, M. S., Pogge, R. W., Stanek, K. Z., ApJ, 695, 259

\bibitem[Perez et al. (2011)]{per11} % gas dependence, MZR
        Perez, J., Michel-Dansac, L., Tissera, P. B.,
	MNRAS, 2011, 417, 580

\bibitem[Perez et al. (2009a)]{per09a} % control
        Perez, M. J., Tissera, P. B., Blaizot, J., 2009, MNRAS, 397, 748

\bibitem[Petric et al. (2011)]{pet11}
        Petric, A. O., et al., 2011, ApJ, 730, 28

\bibitem[Powell et al. (2013)]{pow13}
        Powell, L. C., Bournaud, F., Chapon, D., Teyssier, R., 2013,
	MNRAS, in press

\bibitem[Ramos Almeida et al. (2011)]{ra11}
        Ramos Almeida, C., Tadhunter, C. N., Inskip, K. J., Morganti, R., 
	Holt, J.; Dicken, D.,  2011, MNRAS, 410, 1550

\bibitem[Ramos Almeida et al. (2012)]{ra12}
        Ramos Almeida, C., Bessiere, P. S., Tadhunter, C. N., Perez-Gonzalez,
	P. G., Barro, G., Inskip, K. J., Morganti, R., 
	Holt, J.; Dicken, D.,  2012, MNRAS, 419, 687

\bibitem[Reichard et al. (2009)]{rei09}
         Reichard, T. A., Heckman, T. M., Rudnick, G., Brinchmann, J., 
	 Kauffmann, G., Wild, V., 2009, ApJ, 691, 1005

\bibitem[Robaina \& Bell (2012)]{rb12}
        Robaina, A. R., \& Bell, E. F.,  2012, MNRAS, 427, 901

\bibitem[Rogers et al. (2009)]{rog09}
        Rogers, B., Ferreras, I., Kaviraj, S., Pasquali, A., Sarzi, M.,
	2009, MNRAS, 399, 2172

\bibitem[Rupke, Kewley \& Barnes (2010)]{rkb10} %simulations
        Rupke, D. S. N., Kewley, L. J., Barnes, J. E.,
	 2010, ApJ, 710, L156

\bibitem[Rupke, Kewley \& Chien (2010)]{rkc10} % observations
        Rupke, D. S. N., Kewley, L. J., Chien, L.-H.
	 2010, ApJ, 723, 1255

\bibitem[Sabater et al. (2013)]{sbaf13}
        Sabater, J., Best, P. N., Argudo-Fernandez, M.., 2013,
	MNRAS, 430, 638

\bibitem[Schmidt et al. 2013]{sch13}
        Schmidt, K. B., et al., 2013, MNRAS, 432, 285

\bibitem[Scudder et al. (2012a)]{cg}
         Scudder, J. M., Ellison, S. L., \& Mendel, J. T., 2012a, MNRAS,
	 423, 2690

\bibitem[Scudder et al. (2012b)]{[pairs}
         Scudder, J. M., Ellison, S. L., Torrey, P., Patton, D. R.,
	 Mendel, J. T., 2012b, MNRAS, 426, 549

\bibitem[Simard et al. (2011)]{sim11}
        Simard, L., Mendel, J. T., Patton, D. R., Ellison S. L., 
	McConnachie, A. W., 2011, ApJS, 196, 11

\bibitem[Smith et al. (2007)]{smi07}
        Smith, B. J., Struck, C., Hancock, M., Appleton, P. N., 
	Charmandaris, V., Reach, W. T.,  2007, AJ, 133, 791

\bibitem[Smith et al. (2012)]{smi12}
        Smith, B. J., Swartz, D. A., Miller, O., Burleson, J. A., 
	Nowak, M. A., Struck, C., 2012, AJ, 143, 144

\bibitem[Springel, Di Matteo \& Hernquist (2005)]{sdh05}
        Springel, V., Di Matteo, T., Hernquist, L.,
	2005, MNRAS, 361, 776

\bibitem[Stasinska et al. (2006)]{stas06}
        Stasinska, G., Cid Fernandes, R., Mateus, A., Sodre, L., 
	Asari, N. V.,  2006, MNRAS, 371, 972

\bibitem[Stasinska et al. (2008)]{stas08}
       Stasinska, G., Vale Asari, N., Cid Fernandes, R., Gomes, J. M., 
       Schlickmann, M., Mateus, A., Schoenell, W., Sodre, L., Jr., 
       2008, MNRAS, 391, L29

\bibitem[Stickley \& Canalizo (2013)]{sc13}
        Stickley, N. R., \& Canalizo, G., 2013, ApJ, submitted

\bibitem[Torrey et al. (2012)]{torr2012}
         Torrey, P., Cox, T. J., Kewley, L. J., Hernquist, L.,
	  2012, ApJ, 746, 108

\bibitem[Treister et al. (2012)]{tre12}
         Treister, E., Schawinski, K., Urry, C. M., Simmons, B. D.,
	 2012, ApJ, 758, L39

\bibitem[Wild, Heckman \& Charlot (2010)]{whc10}
        Wild, V., Heckman, T., Charlot, S., 2010, MNRAS, 405, 933

\bibitem[Wong et al. (2011)]{wong11}
         Wong, K. C., et al., 2011, ApJ, 728, 119

\bibitem[Woods \& Geller (2007)]{wg07}
        Woods, D. F., Geller, M. J., 
	2007, AJ, 134, 527

\bibitem[Woods et al. (2006)]{wgb06}
        Woods, D. F., Geller, M. J., Barton, E. J.,
	2006, AJ, 132, 197

\bibitem[Woods et al. (2010)]{woods10}
        Woods, D. F., Geller, M. J., Kurtz, M. J., Westra, E., Fabricant, D.
	 G., Dell'Antonio, I., 2010, AJ, 139, 1857

\bibitem[Xu et al. (2012)]{xu12}
        Xu, C. K., et al., 2012, ApJ, 760, 72

\bibitem[Yan \& Blanton (2012)]{yb12}
        Yan, R., \& Blanton, M. R., 2012, ApJ, 747, 61

\bibitem[Yuan et al. (2010)]{yks10}
        Yuan, T.-T., Kewley, L. J., Sanders, D. B., 2010, ApJ, 709, 884

\end{thebibliography}
\end{document}